\documentclass{article}
\usepackage{spconf,amsmath,amssymb,graphicx,cite}
\usepackage{array}
\usepackage{multirow}
\usepackage{booktabs}
\usepackage{tabularx}
\usepackage{caption}

\title{Harmonics Based Representation in Clarinet Tone Quality Evaluation\\
}
\name{Yixin Wang$^{1}$ \qquad Xiaohong Guan$^{1,2}$ \qquad Youtian Du$^{1,\ast}$ \qquad Nan Nan$^{1}$}

\address{$^{1}$MOE KLINNS Lab, Faculty of Electronics and Information Engineering, \\
Xi’an Jiaotong University, Xi’an, 710049 China.\\
	    $^{2}$Center for Intelligent and Networked Systems, Tsinghua University, Beijing, 100084 China.\\
	    $^{\ast}$Corresponding author. Email: duyt@mail.xjtu.edu.cn}
\begin{document}
\topmargin=0mm
\maketitle
\begin{abstract}
Music tone quality evaluation is generally performed by experts. It could be subjective and short of consistency and fairness as well as time-consuming. In this paper we present a new method for identifying the clarinet reed quality by evaluating tone quality based on the harmonic structure and energy distribution. We first decouple the quality of reed and clarinet pipe based on the acoustic harmonics, and discover that the reed quality is strongly relevant to the even parts of the harmonics. Then we construct a features set consisting of the even harmonic envelope and the energy distribution of harmonics in spectrum. The annotated clarinet audio data are recorded from 3 levels of performers and the tone quality is classified by machine learning. The results show that our new method for identifying low and medium high tones significantly outperforms previous methods.

\end{abstract}
%

%\vspace{-0.5em}
\section{Introduction}
\label{sec:intro}
\vspace{-0.5em}

Persistent practice for many years in controlling breath, fingers, tongue and lips is required to approach the desired tone quality for students to perform a woodwind instrument. Since beginners are often unable to evaluate the tone quality they played, guidance of experts is imperative during practicing. However, the evaluation simply based on the subjective judgement of teachers could be inconsistent as well as time-consuming. Automatic tone quality evaluation of instrumental music has both theoretical and practical values and the techniques of acoustic identification and evaluation can play important roles in speech recognition \cite{Guoe2eSpeechRecognition2019}, speaker verification \cite{GuoSpeakerVerification2018} and music identification \cite{WangRecognizingMusic2018}, etc., as well as music student audition and instrument manufacturing. 

%Years of persistent practice in controlling breath, fingers, tongue and lips is required to approach the desired tone quality for students trying to pick up a woodwind instrument. Since beginners are often unable to distinguish the tone quality they played, the guidance of experts is imperative during this practicing. However, the manual evaluation based simply on the subjective judgement of teachers and experts not only takes a considerable amount of time and energy, but may also contain unfairness and inconsistency in judgment. Automatic tone quality evaluation of instrumental music has great theoretical significance and practical values, as automatic identification technology is widely used in speech recognition \cite{Guoe2eSpeechRecognition2019}, speaker verification \cite{GuoSpeakerVerification2018} and music identification \cite{WangRecognizingMusic2018}, etc. Automatic identification technology can contribute to evaluating both student renditions and the manufacture of clarinet reed.

Many efforts are made to study the quality of music instruments. The efforts were made on influence of physical components \cite{pinardMusicalQualityAssessment2003}, and instrument controlling skills in rendition \cite{faselNeuralNetworkBased1999a,barthetClarinetControlTimbre2010} and tone colors in music context \cite{yangAutomaticTimbreChineseFlute2007,yoritaUsingSpectralAnalysis2015}. The studies on the evaluation of saxophone timbre were considered as lack of systematic understanding of how the features contribute to tone quality \cite{yu-hsianghsiaoMulticlassMTSSaxophone2009,GuoTimbreSaxophone2015,leiToneQualityRecognition2016}. In \cite{chavezASSESSINGQUALITYSOUND2015}, Chavez et al. showed that the clarinet tone quality is distinguishable visually via spectrogram, and in \cite{laAnalyzingQualityClarinet2017} the image representation by a popular convolutional neural network (CNN) structure, AlexNet, is analyzed with an average accuracy of 76.56\%. Yet, few have analyzed the tone quality with respect to different quality of reeds and found an effective representation of tone quality.

%Researchers have approached from several perspectives to study the quality of instruments. Some focused on the influence of physical components \cite{pinardMusicalQualityAssessment2003}, and others paid attention to the instrument controlling skills in rendition \cite{faselNeuralNetworkBased1999a,barthetClarinetControlTimbre2010} and tone colors in music context \cite{yangAutomaticTimbreChineseFlute2007,yoritaUsingSpectralAnalysis2015}. Previous studies on the evaluation of saxophone timbre have been impeded by the lack of a systematic understanding of how the features contribute to tone quality \cite{yu-hsianghsiaoMulticlassMTSSaxophone2009,GuoTimbreSaxophone2015,leiToneQualityRecognition2016}. In \cite{chavezASSESSINGQUALITYSOUND2015}, Chavez et al. showed that the clarinet tone quality is distinguishable visually via spectrogram, and in \cite{laAnalyzingQualityClarinet2017}, they analyzed this image representation by a popular convolutional neural network (CNN) structure, AlexNet, reaching an average accuracy of 76.56\%. Yet, few have analyzed the tone quality with respect to different quality of reeds and found an effective representation of tone quality which contributes to the best tone quality classification.

Differentiating the quality of reed and that of pipe is very important for selecting reed and improving pipe quality of clarinet and other woodwind instruments with reed. In this paper, we present a new method for identifying the clarinet reed quality by evaluating tone quality. By analyzing the clarinet acoustic model and the clarinet audio signal we discover that the tone quality could be decoupled to that of reed and pipe in terms of harmonics. As pipe mainly produce the odd harmonics, the clarinet reed quality depends largely on the even parts of the harmonic series. The harmonic energy distribution in spectrum is highly related to tone quality and harmonic-to-noise ratios in different spectrum ranges is utilized as a part of tone representation. Thus, the tone quality representation is based on two key parts: the harmonic structure and harmonic energy distribution. 

The annotated clarinet audio data are recorded from 3 levels of performers  and a support vector machine (SVM) based classifier is developed. The effectiveness of the representation proposed in the paper is demonstrated by the performance data with two classification strategies: single-note strategy and multi-notes strategy. The results show that our method significantly outperforms previous methods.
\section{RELATION TO PRIOR WORK}
\label{sec:prior}
\vspace{-0.5em}

Our efforts focus on the representation of clarinet tone and its quality evaluation related to instrument physics, acoustic signal analysis and music theory. Most related work carried out by previous researchers focused on the classification techniques, such as random forest in multi-class classification \cite{leiToneQualityRecognition2016} and the structures of CNN \cite{laAnalyzingQualityClarinet2017}. A few efforts were made on the feature representation of instrumental tone quality. Hsiao et al. \cite{yu-hsianghsiaoMulticlassMTSSaxophone2009} analyzed the waveform-shape-based features of saxophone, whereas systematical investigation is not conducted on the belief that tone quality relies on the richness of harmonics in frequency domain. Guo et al. \cite{GuoTimbreSaxophone2015} constructed a one-dimension feature from energy distribution but without clarifying the relation to timbre. In addition, researchers have attempted to evaluate physical structures in a clarinet using optical holography \cite{pinardMusicalQualityAssessment2003} and investigated instrumental performance techniques \cite{faselNeuralNetworkBased1999a,barthetClarinetControlTimbre2010} as well as the timbre of instruments such as bright, sweet, thick and transparent \cite{yangAutomaticTimbreChineseFlute2007,yoritaUsingSpectralAnalysis2015}.
%Our work focuses on the representation of clarinet tone and its quality evaluation, which takes advantage of the integration of the instrument physics, music theory and signal analysis. Most related work carried out by other authors put more emphasis on the classification techniques, such as random forest in multi-class classification \cite{leiToneQualityRecognition2016} and the structures of CNN \cite{laAnalyzingQualityClarinet2017}, rather than an effective representation of tone quality. Few works studied the feature representation of instrumental tone quality. Hsiao et al. \cite{yu-hsianghsiaoMulticlassMTSSaxophone2009} analyzed the waveform-shape-based features of saxophone, whereas they had yet to systematically investigated the common sense in music field that the tone quality relies on the richness of harmonics in frequency domain. Guo et al. \cite{GuoTimbreSaxophone2015} constructed a one-dimension feature but did not clarify the relation to timbre. %???????????????In addition, researchers have attempted to evaluate physical structures in producing sound in a clarinet using optical holography \cite{pinardMusicalQualityAssessment2003}.  A few have investigated instrumental performance techniques \cite{faselNeuralNetworkBased1999a,barthetClarinetControlTimbre2010} as well as the timbre of instruments such as bright, sweet, thick and transparent. \cite{yangAutomaticTimbreChineseFlute2007,yoritaUsingSpectralAnalysis2015}.

Most of these studies have not addressed the identification of reed quality. Our study focuses on the instrument physical model and its acoustic signal with insight into the representation of clarinet tone quality.
%Most of these studies have suffered from a lack of well-grounded theoretical considerations in sound generation and not addressed the identification of reed quality, the sound source. Our study involves the analysis of instrument physical models and acoustic signals, and it capitalizes on a deeper insight into the representation of clarinet tone quality, which was not considered in earlier studies.

% \pagebreak  %%%%%%%%%%%%%%%%%%%%%%%%%%%%%%%%%%%%%%%%%%%%%%%%%%%%%%%%%%%%%%%%%%%%%
%\vspace{-0.5em}
\section{Harmonics based model and clarinet tone evaluation method}
\label{sec:methods}
\vspace{-0.5em}

\subsection{Harmonic Structure Features}
\label{ssec:HSF}
\captionsetup[figure]{singlelinecheck=off,justification=raggedright}
\begin{figure}[htb]
\setlength{\abovecaptionskip}{1pt}
\setlength{\belowcaptionskip}{0pt}
\begin{minipage}[b]{1.0\linewidth}
  \centering
  \centerline{\includegraphics[width=8.5cm]{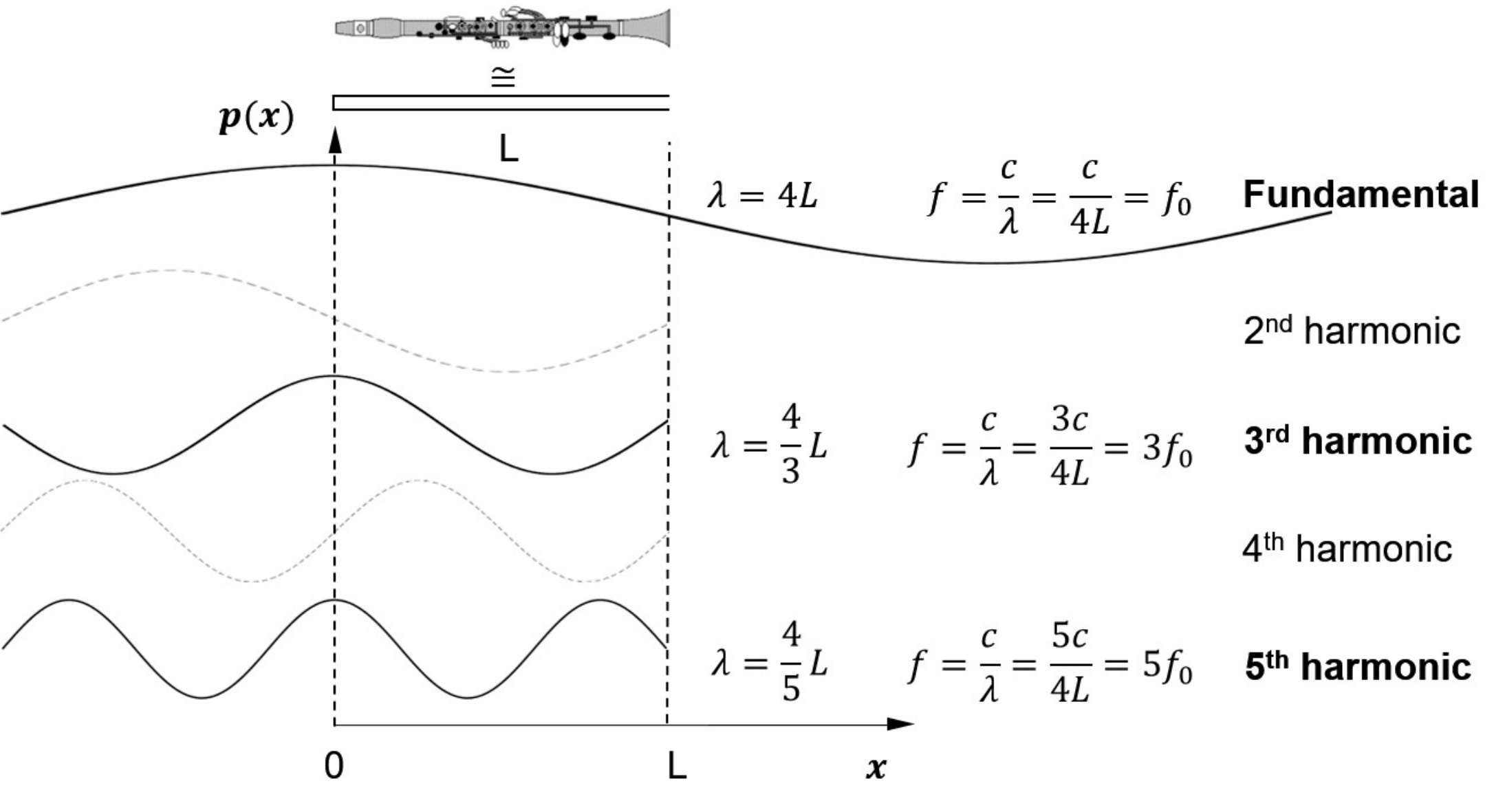}}
%  \vspace{1.5cm}
\end{minipage}
\caption{Clarinet plays (approximately) odd members of the harmonic series only}
\label{fig:harmonic}
\end{figure}

An instrument can be formulated as an input-output system model. For a clarinet, the single vibrating reed, pipe and output signal are considered as the input, system and output of the model, respectively.
As shown in Fig.\ref{fig:harmonic}, the clarinet is modeled as a cylindrical pipe of length $L$ opening at the far end (i.e., bound-unbound boundary) but almost closed at the other end (i.e., hard boundary) because the aperture between reed and mouthpiece is tiny enough to cause a reflection almost like that from a completely closed end \cite{howardAcousticsPsychoacoustics2009}. 
We distinguish the open end and closed end by coordinates $x = L$ and $x = 0$, respectively.
The acoustic pressure at a point at $x \in [0, L]$ can be described by the ideal acoustic model in Eq.\ref{eq:acfun}, which is the superposition of the bi-directional propagation of waves in the pipe \cite{rossingPrinciplesVibrationSound2010}.
\begin{equation}
    p(x,t) = (A e^{-j k x}+B e^{j k x})e^{j\omega t}  \label{eq:acfun}
\end{equation}
where $A$ and $B$ are the amplitudes of the bi-directional waves, $\omega=2\pi f$ is the angular frequency, $k=\frac{2\pi}{\lambda}=\frac{\omega}{c}$ is the angular wave number, and $c$ is the speed of sound in air. 
In the acoustic model, the pressure node and pressure antinode appear at the bound-unbound boundary and hard boundary, accordingly (i.e., $p(x=L)=0$ and $p(x=0)=p_{max}$).
Consequently, only the standing waves with odd numbers of quarter wavelength are allowed between the two boundaries, and the frequencies can be expressed by:
\begin{equation}
f_{n} = \frac{(2n+1)c}{4L},\; n = 0, 1, 2, ...  \label{eq:oddharmof}
\end{equation}
where $f_{n}$ is the frequency of the $n$-th standing wave.
% As we known that a tone consists a variety of frequency components, the  smallest sound frequency $f_0 = \frac{c}{4L}$ supported by the pipe is called the fundamental frequency ($F0$) which determines the lowest pitch and those frequencies that are integer multiples of the fundamental frequency are called harmonics \cite{MathematicsMusic}.  (analog-to-digital converter)
The sound is described as acoustic pressure in physics, and is measured with an electrical transducer. The sampled signal $s(n)$ is defined as \cite{hjerrildEstimationGuitarString2019}:
\begin{equation}
s(n) \ {\propto} \ p(x,t)  \label{eq:xp}
\end{equation}
Actually, the captured sound $s(n)$ is a superposition of the sound generated directly by the reed and the pipe:
\begin{equation}
    s(n)=s_{reed}(n) + s_{pipe}(n)  \label{eq:fun_reedpipe}
\end{equation}

It is demonstrated that the sound quality has a strong correlation with harmonics \cite{kreimanUnifiedTheoryVoice2014, garellekModelingVoiceSource2016}. 
Here, we can rewrite the signal $x(n)$ into a Fourier series with a harmonic part and a noise part. 
The signal produced by the reed and pipe can be rewritten as follows:
\begin{equation}
    s_{reed}(n) = \sum_{m=0}^{M} \beta_m e^{j\omega_0 m n} + \nu_2(n)
    \label{eq:fun_pipe}
\end{equation}
\begin{equation}
    s_{pipe}(n) = \sum_{m=0}^{M} \alpha_m e^{j\omega_0 (2m+1) n} + \nu_1(n)
    \label{eq:fun_reed}
\end{equation}
where $m$ is the index of the harmonics, $\omega_0$ is the fundamental angular frequency, $\alpha_m$ and $\beta_m$ are constant coefficients, and $\nu_1(n)$ and $\nu_2(n)$ are the random noise. As analyzed above, the clarinet pipe produces only the odd harmonics.

%We proposed to parametrize $x(n)$ with an ideal harmonic model, as many researchers have studied that the sound quality has a strong correlation with its harmonics \cite{kreimanUnifiedTheoryVoice2014, garellekModelingVoiceSource2016}. 
% At time instance $n$, the observed complex-valued signal vector $\mathbf{x}$ is represented as $\mathbf{x} = [x_0 \ x_1 \ x_2 \ ... \ x_{N-1}]^T$. A complex signal can ease both notation and computational complexity and a real-valued signal is converted to complex signal by using the Hilbert transform. \cite{marpleComputingDiscretetimeAnalytic1999}
%where $m$ is the number of harmonics, $\omega_0$ is the fundamental angular frequency, $\alpha_m$ and $\beta_m$ are constant coefficients, $\nu_1(n)$ and $\nu_2(n)$ are the random noise. The output signal is attributed to the comprehensive vibration of reed and pipe.
%\begin{equation}
%    x(n)=x_{reed}(n) + x_{pipe}(n)  \label{eq:fun_reedpipe}
%\end{equation}

By substituting Eq.\ref{eq:fun_pipe} and Eq.\ref{eq:fun_reed} into Eq.\ref{eq:fun_reedpipe}, we can obtain:
\begin{equation}
\begin{split}
    s(n)&= \sum_{m=0}^{M} [\gamma_m e^{j\omega_0 (2m+1) n} 
    + \beta_m e^{2j \omega_0 m n}] + \nu(n)
    \\&= s_{odd} + s_{even} + \nu(n)  \label{eq:fun_oddeven}
\end{split}   
\end{equation}
where $\nu(n)=\nu_1(n)+\nu_2(n)$, \; $\gamma_m = \alpha_m+\beta_m$, the fundamental angular frequency $\omega_0 = 2\pi f_{0} = \frac{\pi c}{2L}\label{eq:oddharmow}$.
We note that the odd harmonics are the combination of signals coming from both reed and pipe, whereas the even harmonics depend on reed vibration only. 
We also observe in experiments that the amplitudes of odd harmonics $s_{odd}$ are larger than those of even harmonics $s_{even}$ in lower frequencies, which has also been reported in \cite{howardAcousticsPsychoacoustics2009}.
Thus, we consider that the even harmonics encode the characteristics of reeds and are more effective in evaluating their quality. 

% Acoustically, a note perceived to have a distinct pitch contains frequency components that are integer multiples of $F0$ usually known as harmonics”. Each harmonic is a sine wave and since the hearing system analyzes sounds in terms of their frequency components it turns out to be highly instructive, in terms of understanding how to analyze and synthesize periodic sounds, as well as being central to the development of Western musical harmony, to consider the musical relationship between the individual harmonics themselves.

% The musical intervals of adjacent harmonics in the natural harmonic series starting with the first harmonic, illustrated on a musical stave are: octave (2:1), perfect fifth (3:2), perfect fourth (4:3), major third (5:4), minor third (6:5) \cite{hindemith1941craft}.
% The development of Western harmony follows a pattern where the intervals central to musical development have been gradually ascending the natural harmonic series. These changes have occurred partly as a function of increasing acceptance of intervals which are deemed to be musically consonant, or pleasing to listen to, as opposed to dissonant, or unpleasant to the listener.

%Furthermore, the musical intervals in harmonics series also have an effect on the evaluation of reed quality.
%In the musical stave, there are various intervals of adjacent harmonics starting with the fundamental frequency, such as octave (2:1), perfect fifth (3:2), perfect fourth (4:3), major third (5:4), minor third (6:5) \cite{hindemith1941craft}. 
Furthermore, as experimentally analyzed and concluded in \cite{plompTonalConsonanceCritical1965,berezovskystructure2019}, the intervals, namely, unison, octave, perfect fifth and perfect fourth are pleasing to listen and are called perfect consonances. 
% Major thirds, minor sixths, minor thirds and major sixths are less pleasant to the listener and are called imperfect consonances. The others intervals are unpleasant and are called dissonances.
Since the first four harmonics constitute a perfect consonance with adjacent intervals of octave, perfect fifth and perfect fourth, their relative amplitudes in dB are strongly related to the nuances of tone quality.
%to determine to what extent two sine waves played together sound consonant, or pleasing to listen to, as opposed to dissonant, or unpleasant to the listener as their frequency difference is altered \cite{plompTonalConsonanceCritical1965}. 
Therefore, we extract the following harmonics-based features for the clarinet tone quality representation from the even parts of the first four harmonics and an overall spectral shape described by harmonics:
%Therefore, we studied the first four harmonics since their intervals are perfect consonance and extracted the following harmonics-based features produced by reeds of good, normal and bad quality as clarinet timbre representation,
%we denote the feature set with $\theta$, i.e.
\begin{equation}
    \begin{split}
        HSF = \{H_2, H_4, H_1-H_2,H_2-H_4,\\H_4-H_{i_{2k}},H_{i_{2k}}-H_{i_{5k}}\} \nonumber
    \end{split}
\end{equation}
where $HSF$ denotes the harmonic structure feature set, $H_1$, $H_2$ and $H_4$ are the amplitude of the first, second and fourth harmonics, $H_{i_{2k}}$, $H_{i_{5k}}$ are the harmonics nearest to 2kHz and 5kHz, commensurately, and $H_1-H_2,H_2-H_4,H_4-H_{i_{2k}},H_{i_{2k}}-H_{i_{5k}}$ are the differences between two harmonic amplitudes.
From the analysis above, we consider $H_2$ and $H_4$ in this feature set as a result of reed vibration.
The differences between harmonics are inspired by a psycho-acoustic model of voice quality and have effectively been applied to speaker verification in \cite{garellekModelingVoiceSource2016}.
In our case, $H_1-H_2$, $H_2-H_4$ contain important information of sound quality: the spectral slope of odd-to-even harmonic and even-to-even harmonic.
The high frequency components, $H_4-H_{i_{2k}}$ and $H_{i_{2k}}-H_{i_{5k}}$ indicate the spectral noise level and are negatively correlated with the quality of sound. In general, a higher level of high frequency noise leads to a lower level of tone quality. 

\vspace{-0.5em}
\subsection{Harmonic Energy Features} %==============================================================================\vspace{-1em}
\label{ssec:HEF}

Energy is a measure of amount of sound produced by reed vibration. Previous studies \cite{pinardMusicalQualityAssessment2003,GuoTimbreSaxophone2015} present that the distribution of energy and symmetry of reed in vibration accounts for tone quality. 
The symmetry of reed in vibration results in periodicity in signal, so we consider a representation associated with harmonics and noise, indicating periodic and non-periodic parts separately.
Thus the harmonic-to-noise ratio (HNR) is calculated as follows:
\begin{equation}
\begin{split}
    HNR = \frac{ \sum_{n=0}^{N-1} h^2(n)}{\sum_{n=0}^{N-1} v^2(n)}
\end{split}   
\end{equation}
where $h(n)$ is the harmonic component, $v(n)$ is the noise component and the signal $s(n)$ is a mixture of them $s(n)=h(n)+v(n)$. $N$ is the number of samples in a spectral segment.
In order to describe the energy distribution of harmonics, we divided the spectrum into five parts, containing fundamental and harmonics in lower frequency.
% : 0-500Hz, 500-1500Hz, 1500-2500Hz, 2500-3500Hz and the frequencies above 3500Hz. 
As we are more interested in the lower frequency parts, we construct the feature as follows: 
\begin{equation}
    \begin{split}
        HEF = \{HNR_{05}, HNR_{15}, HNR_{25}, HNR_{35}, RMSE\} \nonumber
    \end{split}
\end{equation} 
where $HEF$ denotes the harmonic energy feature set, $HNR_{05}$, $HNR_{15}$, $HNR_{25}$, $HNR_{35}$ are the harmonic-to-noise ratios within the frequencies of 500Hz, 1500Hz, 2500Hz and 3500Hz. $RMSE$ is the root mean square of total energy. A better tone quality attributes to a higher HNR and a lower overall energy.

%  \vspace{-1em}
\subsection{Clarinet Tone Quality Classification} %---------------------------------------
\label{ssec:classification}

%\vspace{-1em}

\begin{figure}[htb]
\setlength{\abovecaptionskip}{1pt}
\setlength{\belowcaptionskip}{0pt}
\begin{minipage}[b]{1.0\linewidth}
  \centering
  \centerline{\includegraphics[width=8.5cm]{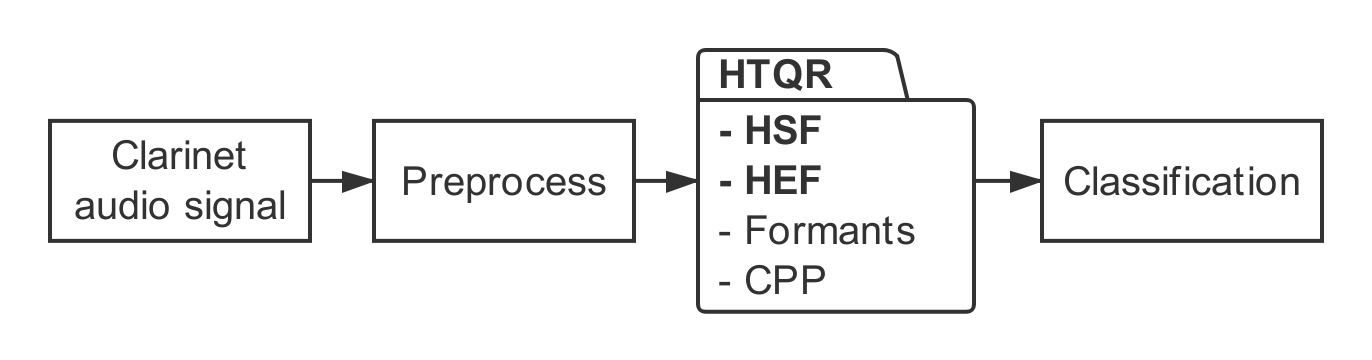}}
%  \vspace{1.5cm}
\end{minipage}
\caption{Scheme of clarinet tone quality classification.}
\label{fig:system}
\end{figure}
\vspace{-1em}
As shown in Fig.\ref{fig:system}, the scheme of clarinet tone quality classification is composed of preprocessing, feature extraction and classification.
Clarinet audio signals are sampled at the rate of 96kHz and we intercept the steady state by removing the transient state segment and decay state segment.
We construct a $d_{f}$-dimensional feature vector, named harmonics-based tone quality representation (HTQR), to represent clarinet audio signals, where the harmonic structure features (HSF) and the harmonic energy features (HEF) introduced above are adopted as key features. Other complementary audio features are the frequencies and bandwidths of the first four formants (Formants) and cepstral peak prominence (CPP).
%The audio feature set used for classification denoted as HTQR (Harmonics-based Tone Quality Representation) is comprised of the harmonics-based features and some frequently used audio features such as fundamental frequency(F0), cepstral peak prominence(CPP), signal to noise ratio(SNR) and the frequencies and bandwidths of the first four formants. 
Features in HTQR are achieved after a short time Fourier transform performed using a sliding window of width 25ms and step size 10ms with Voice Sauce software \cite{ShueVoiceSauce2009}.
%The first twenty Mel Frequency Cepstral Coefficients(MFCCs) were also calculated as a baseline which were proved to be effective in the tone quality discrimination in \cite{loganMelFrequencyCepstral2000,leiToneQualityRecognition2016}. 
The SVM classifier using the radial basis function (RBF) kernel is utilized to evaluate the performance of the features. Considering the dataset is not large, more complex classifiers may outfit. 
In the work, we divide the tone quality into three levels and implement the classifiers with one-versus-one strategy to address the multi-class classification problem.
%Penalty in the error term was also considered with parameter $C = 1$ to improve the generalization ability of the system.

%with parameter $\gamma = \frac{1}{d_{f}}$, where $d_{f}$ is the dimension of feature set, 

%\vspace{-0.5em}
\section{Experimental Results}   %%%%%%%%%%%%%%%%%%%%%%%%%%%%%%%%%%%%%%%%%%%%%%%%%%%%%%%%%%%%%%%%%%
\label{sec:exp}
\vspace{-0.5em}
\subsection{The Experimental Testbed} %-----------------------------------------
\label{ssec:setup}
\vspace{-0.5em}
A clarinet tone dataset is created in a professional recording room using professional equipment: a condenser microphone, a voice channel microphone preamp and an audio interface. 
%For the evaluation, we construct a clarinet tone dataset, which was directly recorded in a professional recording room using professional equipment such as a condenser microphone, a voice channel microphone preamp and an audio interface. 
Three professional clarinet performers from Xi'an Conservatory of Music are invited for data collection.
%The dataset used in this study was created especially for clarinet tone analysis because it is ideal to use the recording of real instrument without polishing.
We utilize a B$\flat$ clarinet and 80 reeds of three levels of quality: low level, medium level and high level. For data collection, we choose 13 tones from the frequently used range (E3 to E6) covering the chalumeau, clarion and altissimo register of the B$\flat$ clarinet. For each tone, every reed is used to generate nine 3-second samples. 
In total, the dataset consists of $9\,360$ audio samples and lasts around 7.8 hours. There are $2\,340$, $3\,510$ and $3\,510$ samples at the low, medium and high level of quality, correspondingly.

The data is randomly split into two parts: 75\% for training and 25\% for testing. We obtain 20-dimensional HTQR feature vector (i.e., $d_f = 20$) for feature extraction. In the classification process, the kernel width of RBF is set to $\gamma = \frac{1}{d_{f}}$, and penalty in the error term is considered with parameter $C = 1$ to improve the generalization ability.
%of the system.

To study the dependency of classification results on different tones, we design the experiment with the following two strategies: multi-notes strategy (MN) builds a single classification model for the samples of all 3 different qualities and 13 different tones; single-note strategy (SN) builds multiple classification models, each for the samples of the same tone. Four single-note datasets are used in the experiment, namely E3 (165Hz), E4 (330Hz), E5 (659Hz), and E6 (1319Hz).
%\begin{itemize}
%  \item Multi-notes strategy (MN): building a single classification model for the samples of all 3 different qualities and 13 different tones;
%  \item Single-note strategy (SN): building multiple classification models, each for the samples of the same tone. Four single-note datasets are used in the experiment, namely E3 (165Hz), E4 (330Hz), E5 (659Hz), and E6 (1319Hz).
%\end{itemize}

%After feature extraction, we evaluate both MFCC and the proposed HTQR with SVM classifier.
%Then we performed an ablation experiments to show that the components are necessary features concerning the representation of tone quality without redundancy.

\vspace{-0.3em}
\subsection{Ablation Study for HTQR}
\label{ssec:ablation}
\vspace{-0.3em}
In this experiment, we examine how variation in the proposed feature set affects tone quality evaluation and show the effectiveness of HTQR for different tones. 
We test five kinds of representation in the ablation study, where HTQR is the complete data set, $\text{HTQR}\setminus\text{HSF}$ corresponds to the features generated by removing HSF from HTQR, and the other cases are presented in a similar manner. 
%: HTQR, $\text{HTQR}\setminus\text{HSF}$, $\text{HTQR}\setminus\text{HEF}$, $\text{HTQR}\setminus\text{Formants}$ and $\text{HTQR}\setminus\text{CPP}$
The performance was evaluated by F1-score and accuracy metrics as appropriate.
%By testing the results of the classification after each ablation, we could determine the significance of each ablated feature set.
%The first variation omits the HBF containing even harmonics in low frequencies and their differences. 
%The second ablates the frequencies and bandwidths of the first four formants. 
%The other ablation selections are the fundamental frequency, the signal to noise ratio, the root mean square (RMS) energy and the cepstral peak prominence.  

%\begin{equation}
%\text{F1} = 2 *\frac{\text{precision} * \text{recall}}{\text{precision} + %\text{recall}}  \nonumber \label{eq:F1}
%\end{equation}
% \begin{equation}
% Accuracy = \frac{correctly \ classified \ samples}{all \ examples}\label{eq:accuracy}
% \end{equation}

%As each ablation group of features is removed, the F1-score in Eq.\eqref{eq:F1} and classification accuracy without that group is calculated in the 

Table \ref{tb:ablation} reports the results of ablation study for the five cases. 
HTQR including our proposed HSF, HEF and the other complementary features introduced in section \ref{ssec:classification} achieves the best performance in tone quality evaluation. 
We also observe that $\text{HTQR}\setminus\text{HSF}$ obtains the lowest value in all the cases and  $\text{HTQR}\setminus\text{HEF}$ also largely contributes, which demonstrates that harmonic structure and energy distribution play a crucial role.
In addition, formants indicate the first four harmonics whose intervals are consonant and CPP is an important feature in defect detection.
%in all the features consisted of in HTQR.provide information of
%; SNR and RMSE indicate the extent of the noise level in an audio; F0 carries the pitch information. The latter 5 features contributes to tone quality evaluation as a complementary.
%\vspace{-0.5em}
\captionsetup[table]{singlelinecheck=off,justification=raggedright}
\begin{table}[ht]
\setlength{\abovecaptionskip}{0pt}
\setlength{\belowcaptionskip}{3pt}
\caption{\label{tb:ablation} The comparison results in ablation study.}
\centering
\begin{tabular}{p{4cm}<{\centering}|p{1.5cm}<{\centering} p{1.5cm}<{\centering}}
\toprule[2pt]
\textbf{Features} & \textbf{F1-score} & \textbf{Accuracy}\\
% \hline
\toprule[1pt]
\textbf{HTQR} & \textbf{0.85} & \textbf{0.84}\\
$\text{HTQR}\setminus\text{HSF}$ & 0.77 & 0.78\\
$\text{HTQR}\setminus\text{HEF}$ & 0.79 & 0.79\\
$\text{HTQR}\setminus\text{Formants}$ & 0.81 & 0.80\\
$\text{HTQR}\setminus\text{CPP}$ & 0.84 & 0.85\\
\toprule[2pt]
\end{tabular}
\end{table}

\vspace{-1em}
\subsection{Classification Performance Evaluation on the Multi-notes and Single-note dataset} 
\label{ssec:multi-exp}
\vspace{-0.3em}
In this subsection, we test the performance of HTQR for the tone quality evaluation and compare it with the widely used Mel Frequency Cepstral Coefficients (MFCC) (20-dimensional) \cite{leiToneQualityRecognition2016,loganMelFrequencyCepstral2000}. 
We implement the experiments with two strategies introduced in section \ref{ssec:setup}: MS and SN. 
Table \ref{tb:results} compares the results of HQTR and MFCC with two different strategies. 
We first observe that HTQR achieves the satisfactory performance, i.e., 0.84 and 0.92 in accuracy, with both strategies. Additionally, SN outperforms MN since it results in a lower within-class variance than the latter.
Turning to Fig.\ref{fig:bar}, it provides a more detailed result for the quality evaluation of each tone in the SN strategy.
It depicts that the low-frequency tones, such as E3, E4 and E5 reported significantly a higher F1-score than the high-frequency tones. 
This decline in high frequency appears as well using MFCC.
%We show a more detailed result for the quality evaluation of each tone in the SN strategy in Fig.\ref{fig:bar}. It is illustrated that for the low-frequency tones, such as E3, E4 and E5, we obtain xx F1-score xx, while the performance decreases for high-frequency tones. 
A challenge for both features is to improve performance in high frequency, because the high-frequency tones are sparse in spectrum, that is, carrying less information and easier to be disturbed by high-frequency noises. 
Table \ref{tb:results} also illustrates that the proposed HTQR outstrips the MFCC features with an absolute improvement of 14\% to 20\%. 
%Despite that HTQR might be less efficient in high frequency due to random noise, the experimental results indicate that HTQR is able to identify a nuances in tone quality representation.
%Multi-notes strategy demonstrated that HTQR achieved a performance with accuracy of 87\%, while MFCC only showed the capability to distinguish tone quality of different clarinet reeds with accuracy of 64\%.
%We observed in Fig.\ref{fig:bar} that the comparison between notes has less significant differences in lower frequency, and the score in high frequency is not as good as the performance in lower frequency.
%We highlight that the single-note strategy gives the best result. In this strategy, we selected four single-note datasets, namely E3 (165Hz), E4 (330Hz), E5 (659Hz), and E6 (1319Hz). Since all the samples are of the same pitch, the classifier focuses only on the differences in tone quality within the same note. Thus, single-note strategy achieved a lower within-class divergence than the multi-notes and result in a better classification performance.
%We observed in Fig.\ref{fig:bar} that the comparison between notes has less significant differences in lower frequency, and the score in high frequency is not as good as the performance in lower frequency. 
%By comparing the two strategies, we can conclude that HTQR is quasi-independent to pitch in tone quality identification and improves the accuracy significantly comparing to previous work with both the single-note and multi-note data.
\vspace{-0.25em}
\begin{table}[ht]
\setlength{\abovecaptionskip}{0pt}
\setlength{\belowcaptionskip}{0pt}
\caption{\label{tb:results} The compared results of HTQR and MFCC with two strategies.}
\centering
\begin{tabularx}{8.1cm}{c|c|c|c|cc}
% \hline
\toprule[2pt]
\multirow{2}{*}{\textbf{Feature set}} & \multirow{2}{*}{\textbf{Level}} & \multicolumn{2}{c|}{\textbf{MN}} & \multicolumn{2}{c}{\textbf{SN}(E3)} \\ \cline{3-6} 
 &  & F1 & ACC & \multicolumn{1}{c|}{F1} & ACC \\ 
\toprule[1pt]
\multirow{3}{*}{HTQR} & good & 0.85 & \multirow{3}{*}{0.84} & \multicolumn{1}{c|}{0.92} & \multirow{3}{*}{0.92} \\
 & medium & 0.83 &  & \multicolumn{1}{c|}{0.91} &  \\
 & bad & 0.86 &  & \multicolumn{1}{c|}{0.95} &  \\ \hline
\multirow{3}{*}{MFCC} & good & 0.66 & \multirow{3}{*}{0.64} & \multicolumn{1}{c|}{0.78} & \multirow{3}{*}{0.78} \\
 & medium & 0.60 &  & \multicolumn{1}{c|}{0.75} &  \\ 
 & bad & 0.64 &  & \multicolumn{1}{c|}{0.81} &  \\
\toprule[2pt]
\end{tabularx}
\end{table}

%\subsection{Single-note Experiment} %----------------------------------------
%\label{ssec:single-exp}
\vspace{-2em}
\begin{figure}[htb]
\setlength{\abovecaptionskip}{1pt}
\setlength{\belowcaptionskip}{0pt}
\begin{minipage}[b]{1.0\linewidth}
  \centering
  \centerline{\includegraphics[width=8.5cm]{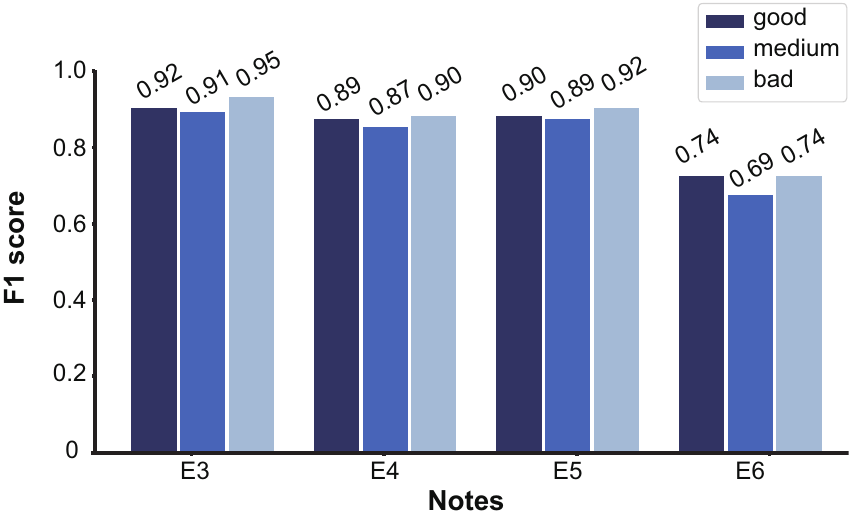}}
%  \vspace{1.5cm}
\end{minipage}
\caption{Performance of HTQR with single-note strategy.}
\label{fig:bar}
\end{figure}

% -------------------------------------------------------------------------
\vspace{-1.6em}
\section{CONCLUSIONS}
\label{sec:conclusions}
\vspace{-0.5em}
In this paper, an effective representation called HTQR is proposed for identification of clarinet reed quality by tone quality evaluation. It is discovered that the quality of reed could be decoupled from that of pipe based on the harmonics of the audio signal.
%We first analyzed the clarinet physical models and acoustic signals and discovered that the signal generated by reed could be separated from pipe.
The pipe mainly produces odd harmonics, and the even harmonics encode the tone characteristics of reeds.
Then based on the reed vibration mode, the harmonic energy with respect to noise is highly relevant to tone quality. 
Accordingly, we extract harmonic structure features (HSF) and harmonic energy features (HEF).
%Then we extracted the feature HSF based on even harmonics and spectral shape concerning tonal consonance.
A feature set HTQR is eventually constructed by combining HSF, HEF and the other 2 widely used features in audio analysis as supplementary. 
%A dataset of clarinet audio was also created with 80 reeds from three levels of quality. 
Testing results based on the dataset demonstrate that HSF play a pivotal role in the tone quality representation of reed. 
We conclude that HTQR significantly improves clarinet tone quality identification with the accuracy of 84\% and 92\% by MN and SN strategy, respectively.
%Further studies regarding extension HTQR to other instruments would be worthwhile.
Our next step is to classify the tone quality of performers with the minimum effect of instrument quality.

%, we plan to adapt this representation to sound quality evaluation for other instruments.
%The method works to classify reeds into three levels, using a recorded clarinet audio dataset from which a harmonics-based feature set was extracted.
%In ablation experiments, we showed that harmonics-based features are pivotal in tone quality identification, and that all the elements in the proposed feature set HTQR are necessary.
%This feature set was sufficient for multi-class SVM classification of tone quality, with a high accuracy for both single-pitch (92.70\%) and multi-pitch data (86.15\%). 
%This method was designed for clarinet audio quality classification, where reeds varied in quality. 
%In future works, we plan to adapt this representation to sound quality evaluation for other instruments.

\vfill\pagebreak

%\section{REFERENCES}
%\label{sec:refs}


\bibliographystyle{IEEEtran} %IEEEbib
\bibliography{refs}

\end{document}